\documentclass[reprint,showpacs,aps,prx,amsmath,amssymb,floatfix,twocolumn]{revtex4-2}

\usepackage{graphicx}
\usepackage[dvipsnames]{xcolor}
\usepackage{siunitx}
\usepackage{MnSymbol}
\usepackage{mathtools}

\newcommand{\bra}[1]{\langle{#1}|}
\newcommand{\ket}[1]{|{#1}\rangle}

\usepackage{natbib}
\usepackage{hyperref}
\hypersetup{
    colorlinks=true,
    citecolor=blue,
    linkcolor=blue,
    anchorcolor=black,
    urlcolor=blue,
}

\usepackage{bm}
\usepackage{xcolor}

\begin{document}
\title{Revealing electron-ytterbium interactions through Rydberg molecular spectroscopy}
\author{Tangi Legrand}
\thanks{\hypertarget{eqcontrib}{These authors contributed equally to this work.}}
\author{Xin Wang\normalfont\textsuperscript{\hyperlink{eqcontrib}{*}}}
\email{{wanxin@iap.uni-bonn.de}}
\author{Florian Pausewang}
\author{Wolfgang Alt}
\author{Eduardo Uruñuela}
\author{Sebastian Hofferberth}
\email{hofferberth@iap.uni-bonn.de}
\affiliation{Institute of Applied Physics, University of Bonn, Wegelerstraße 8, 53115 Bonn, Germany}
\author{Milena Simić$^{1}$}
\thanks{\hypertarget{eqcontrib}{These authors contributed equally to this work.}}
\author{Matthew T.\ Eiles$^{1,2}$}
\affiliation{$^1$Max Planck Institute for the Physics of Complex Systems, Nöthnitzer Straße 38, 01187 Dresden, Germany \\
$^2$Department of Physics and Astronomy, Purdue University, West Lafayette, Indiana 47907, USA}

\date{December 23, 2025}

\begin{abstract}
Divalent atoms have emerged as powerful alternatives to alkalis in ultracold atom platforms, offering unique advantages arising from their two-electron structure. 
Among these species, ytterbium (Yb) is especially promising, yet its anionic properties and its Rydberg spectrum remain comparatively unexplored.
In this work, we perform a first and comprehensive experimental and theoretical investigation of ultralong-range Rydberg molecules (ULRMs) of $^{174}$Yb in $6sns\,^1S_0$ Rydberg states across nearly two decades in principal quantum number $n$ and three orders of magnitude in molecular binding energy.
Using the Coulomb Green’s function formalism, we compute Born–Oppenheimer molecular potentials describing the Rydberg atom in the presence of a ground-state perturber and achieve quantitative agreement with high-resolution molecular spectra.
This enables the extraction of low-energy electron-Yb scattering phase shifts, including the zero-energy $s$-wave scattering length and the positions of two spin-orbit split $p$-wave shape resonances.
Our results provide strong evidence that the Yb$^{-}$ anion exists only as a metastable resonance.%
We additionally show the sensitivity of ULRM spectra to the atomic quantum defects, using this to determine the quantum defect of the $6s23f\, ^1F_3$ state.
Together, these findings establish Yb ULRMs as a powerful probe of electron-Yb interactions and lay essential groundwork for future Rydberg experiments with divalent atoms.
\end{abstract}

\maketitle

\section{Introduction}\label{sec:intro}
Divalent atoms are attracting significant attention in ultracold atomic physics because their electronic structure provides several advantages over the more commonly used alkali atoms~\cite{Shaffer2019, burgdorfer2016, choudhury2024}.
Their spin-forbidden transitions make them ideal for precision metrology, exemplified in their use as optical atomic clocks~\cite{katori2011, schmidt2015, hinkley_atomic_2013, bloom_optical_2014, madjarov_atomic-array_2019}, while their long-lived triplet states expand qubit-encoding possibilities~\cite{zoller2008, endres2020, thompson2022, covey2024}.
Moreover, the presence of a second optically active valence electron enables local coherent control~\cite{cheinet2022, thompson2022, meinert2024a}, cooling, imaging~\cite{burgdorfer2016}, and trapping~\cite{thompson2022, meinert2024}, even when the other electron is excited to a Rydberg state.
Although autoionization poses a challenge in divalent Rydberg atoms and often requires circumvention by choice of Rydberg state~\cite{meinert2024, genevriez2024}, it can also enable powerful tools for control and measurement~\cite{pisharody_probing_2004, millen_spectroscopy_2011, thompson2022, lochead_number-resolved_2013}.
The additional electronic degree of freedom enables further tailoring of Rydberg–Rydberg interactions~\cite{li2024}.
These many features make divalent atoms attractive species for quantum computing and quantum simulation, both in low-lying states~\cite{zoller2008, pagano_fast_2019} and in Rydberg configurations~\cite{endres2020, pagano_error_2022}.

Among divalent atoms, ytterbium (Yb) plays a distinctive role owing to its large mass, wide range of stable isotopes, and experimentally accessible transition energies.
Despite these benefits, it remains less studied than alkali atoms and other divalent species.
This is in part due to the increased complexity of divalent low-$\ell$ Rydberg spectra, which arises from channel mixing with low-lying core-excited states involving the second valence electron and, in Yb, the filled $4f^{14}$ shell.
Although these multi-electron effects can be accurately treated using multichannel quantum defect theory (MQDT)~\cite{Luc-Koenig1996}, the required MQDT parameters cannot currently be calculated \textit{ab initio}, making extensive spectroscopic data essential to constrain the fits~\cite{aymar_highly_1980, lehec2018laser,peper2025spectroscopy, kuroda2025microwave}.
Moreover, even the ground-state properties of Yb remain poorly explored~\cite{andersen2004atomic}.
The static polarizability of Yb has not been calculated to the same precision as in other species, nor has it been measured experimentally~\cite{mitroy2010theory}. 
Whether or not the electron affinity of Yb is negative has been a long-running debate, with the strongest evidence at present suggesting that the electron-atom potential is too shallow to support a bound state~\cite{andersen2004atomic}. 
Such low-energy electron-atom observables are difficult to probe experimentally as photodetachment~\cite{kristiansson2022high,lindahl2012threshold} cannot be employed if a bound anion does not exist, and their sensitive dependence on electron-electron correlation challenges an accurate theoretical description. 

Ultralong-range Rydberg molecules (ULRMs) provide a means by which these properties can be spectroscopically obtained in experiments with ultracold atoms~\cite{sasmannshausen_experimental_2015,engel_observation_2018,peper_heteronuclear_2021, exner_high_2025,giannakeas2020}. 
These weakly-bound molecules form when one or more ground-state atoms are immersed in the highly extended electronic wave function of a Rydberg atom, resulting in bond lengths of several thousand Bohr radii~\cite{sadeghpour2000,hamilton2002shape,chibisov2002energies, pfau2009, rost2010,pfau2010,Shaffer2015,shaffer_ultracold_2018,eiles_trilobites_2019,fey_ultralong-range_2020, dunning_ultralong-range_2024}.
Their binding energies depend sensitively on both the quantum defects of the Rydberg atom and on the low-energy scattering phase shifts of the Rydberg electron off of the ground-state atom, permitting extraction of these parameters in conjunction with an accurate theoretical description of the molecules. 
Such analysis, which has benefited from nearly two decades of intense study of both homo- and heteronuclear ULRMs of Rb, Cs, K, and Sr in ultracold gases~\cite{Shaffer2015, niederprum_observation_2016, deis_observation_2020,desalvo_ultra-long-range_2015}, optical lattices~\cite{manthey_dynamically_2015}, and dual-species tweezer arrays~\cite{guttridge_individual_2025}, is an outgrowth of earlier efforts to extract electron affinities via collisional quenching of Rydberg atoms in gaseous environments~\cite{fabrikant2000quenching,reicherts1997controlled,desfrancois1994prediction,desfranccois19941}. 
ULRMs have established themselves as a versatile tool for exploring few-to-many-body physics and ultracold chemistry~\cite{schlagmuller_ultracold_2016,  srikumar_vibrationally_2025, shaffer_ultracold_2018, fey_ultralong-range_2020, eiles_trilobites_2019, dunning_ultralong-range_2024}, and together with other Rydberg-excited molecular species~\cite{deiss_long_2021,duspayev_long_2020,Pfau2022} compose an increasingly large and fascinating class of long-range molecules. 

In this article, we present a first and extensive experimental and theoretical exploration of ULRM structure in $^{174}\mathrm{Yb}$.
The absence of orbital angular momentum and electron spin in the $^1S_0$ ground state, together with the zero nuclear spin, provides a simple and well-controlled system for studying electron–atom scattering.
We spectroscopically measure ULRM spectra spanning three orders of magnitude in binding energy and almost two decades in principal quantum number. 
To model the system, we employ the Coulomb Green's function formalism to solve for the electronic states in the presence of the ground-state perturber, yielding Born-Oppenheimer molecular potential curves and vibrational spectra which agree with the experimental data to within $\sim\num{10}\%$. 

Our results permit the extraction of Yb$-e^-$ scattering phase shifts, enabling the identification of the zero-energy electron scattering length and the positions of two $p_j$-wave shape resonances in Yb.
Specifically, we provide strong evidence for the conclusion that Yb cannot form an atomic negative ion.%
Additionally, owing to the energy dependence of the $^1F_3$ quantum defect and its close energetic proximity to the $^1S_0$ series, our vibrational molecular spectroscopy provides a highly precise determination of the $6s23f\,^1F_3$ quantum defect, albeit with accuracy limited by model-dependent effects.

\section{Theory of ultralong-range Rydberg molecules}\label{sec:theory}
\begin{figure}
  \centering
  \includegraphics[width=\columnwidth]{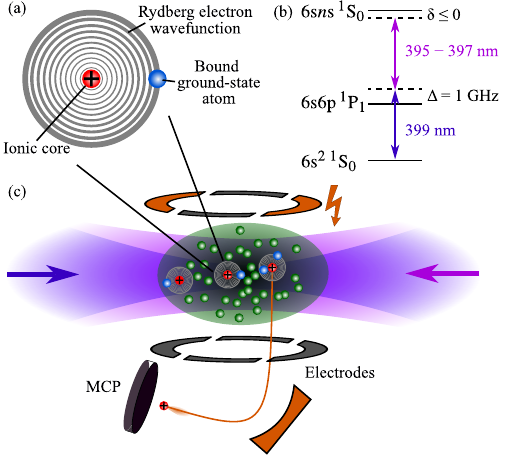}
  \caption{(a)~Schematic drawing of an ultralong-range Rydberg molecule consisting of one (or several) ground-state atom(s) (blue) bound to a Rydberg atom (ionic core in red and electronic wave function of Rydberg electron in gray).
  (b)~Scheme of the photoassociation of Yb Rydberg molecules in the $6sns\,^1S_0$ manifold.
  We use a detuning from the intermediate state $6s6p\,^1P_1$ of $\Delta=\SI{1}{\giga\hertz}$.
  Varying the detuning of the $\SI{399}{\nano\meter}$ laser allows control over the two-photon detuning~$\delta$.
  (c)~Schematic of the experimental setup.
  The pulsed excitation light at $\SI{399}{\nano\meter}$ and at $\SIrange[range-phrase = -]{395}{397}{\nano\meter}$ (depending on the Rydberg state) is focused onto the dense atomic cloud (green).
  The Rydberg molecules or atoms are field ionized after each excitation pulse by high-voltage electrodes (orange) above the atomic cloud.
  The ions are guided towards a microchannel plate (MCP) detector using a deflection electrode.}
  \label{fig:expscheme}
\end{figure}

We begin by reviewing the binding mechanism of ultralong-range Rydberg molecules and discuss how Yb ULRMs provide experimentally accessible insight into both the properties of Yb$^*$ Rydberg atoms and Yb$^-$ anions.

\subsection{Molecular Hamiltonian}
Within the Born-Oppenheimer approximation, the full molecular wave function $\Psi(\vec r,\vec r_c,\vec R)$ factors into a vibrational part $\chi^\alpha_v(\vec R)$ and an electronic part $\phi_\alpha(\vec r,\vec r_c;\vec R)$.
The latter depends parametrically on the internuclear distance $R$, and is an eigenstate of the total electronic Hamiltonian
\begin{equation}
\label{eq:hamiltonian}
\hat H_\mathrm{el}(\vec r,\vec r_c,\vec R) = \hat H_\mathrm{Ryd}(\vec r,\vec r_c) + \hat V_a(\vec r,\vec r_c,\vec R).
\end{equation}
The positions of the two valence electrons relative to the Yb nucleus are denoted by $\vec r$ for the Rydberg electron and $\vec r_c$ for the second valence electron in the ionic core (see Fig.~\ref{fig:expscheme}(a)).
The first term of Eq.~\ref{eq:hamiltonian} is the Hamiltonian of the Rydberg atom, which defines the electronic states $\psi_\alpha(\vec r,\vec r_c)$ via
\begin{equation}\label{eq:rydberghamiltonian}
\begin{aligned}
    \hat H_\mathrm{Ryd}\psi_\alpha(\vec r,\vec r_c)=-\frac{1}{2\nu_\alpha^2}\psi_\alpha(\vec r,\vec r_c),\,\,\,\,\nu_\alpha=n-\mu_\alpha.
\end{aligned}
\end{equation}
The eigenenergies $E_\alpha=-1/2\nu_\alpha^2$ are determined by the integer-valued principal quantum number $n$ and a collection of short-range quantum defect parameters $\mu_\alpha$ which depend on the electronic configuration, generically labeled as $\alpha$.
Throughout, we use atomic units (a.u.) except where specified. 
The effective non-integer principal quantum number is denoted $\nu_\alpha$. 

The second term in Eq.~\ref{eq:hamiltonian} is the interaction of the Rydberg electron with the ground-state Yb atom. 
As first derived by Fermi~\cite{Fermi1934} and later generalized by Omont and others~\cite{omont1977theory,idziaszek2006pseudopotential}, a powerful pseudopotential for $\hat V_a(\vec r,\vec r_c,\vec R)$ is obtained from a parameterization in terms of the energy-dependent $e-$Yb scattering phase shifts $\delta_{L_a}(k)$. 
Here, $L_a$ is the scattering partial wave (we use $a$ to denote quantum numbers referencing the \textit{anionic} system, i.e., the electron-atom complex),  $k=\sqrt{2E+2/R}$ is the electron's momentum, and $E$ is the total energy of the electron in the Coulomb field. 
Because the scattering phase shifts diminish rapidly with increasing $L_a$ as a consequence of the Wigner threshold law, it usually suffices to truncate this expansion at $L_a= 1$~\cite{giannakeas2020}, yielding the pseudopotential
\begin{equation}\label{eq:fermi}
\begin{aligned}
    \hat V_a(\vec r,\vec r_c,\vec R) &= 2\pi a_s(k)\delta^3(\vec r - \vec R) \\&+6\pi a_p^3(k)\overleftarrow{\nabla}\delta^3(\vec r - \vec R)\overrightarrow\nabla.
\end{aligned}
\end{equation}
Here, $a_s(k)=-\tan\delta_s(k)/k$ and $a_p^3(k)=-\tan\delta_p(k)/k^3$ are the energy-dependent $s$-wave scattering length and $p$-wave scattering volume, respectively.

In this study we consider Rydberg states well below the first ionization threshold. 
These are energetically well-separated from doubly excited states where both electrons could scatter off of the distant ground-state atom. 
For this reason, $\hat V_a$ does not explicitly depend on $\vec r_c$.%
The molecular system is effectively a three-body system $\mathrm{Yb}^++ e^- + \mathrm{Yb}$ whose description relies on accurate knowledge of the individual two-body complexes of the neutral Rydberg atom Yb$^*$ and the anionic system Yb$^-$.
These are each described in terms of a set of short-range parameters.
In the former case, these are the quantum defects $\mu_\alpha$  describing the electron's interaction with the Yb$^+$ core and with the second valence electron, while in the latter they are the low-energy scattering phase shifts $\delta_{L_\alpha}(k)$ characterizing the interaction between the neutral Yb atom and the electron. 
When these parameters are known, we solve for the $R$-dependent eigenenergies $U_\alpha(R)$ of the Hamiltonian of Eq.~\ref{eq:hamiltonian}. These form the set of potential energy curves in which the vibrational states $\chi^\alpha_v(\vec R)$ are solved. 

\subsection{Ytterbium anion Yb$^-$}\label{sec:ybanion}

The properties of the $\mathrm{Yb}+e^-$ collisional complex determine whether or not a long-lived $\mathrm{Yb}^*+\mathrm{Yb}$ ULRM can form. 
Little is known about the $s$-wave scattering properties of electrons from Yb atoms, except that the zero-energy scattering length $a_s(0)$ is most likely negative due to the absence of a bound Yb$^-$ state of this symmetry. 
Although some species of similar electronic structure, for example Hg, do have positive $a_s(0)$~\cite{mceachran2003momentum}, the majority -- for example, all alkaline earth metals -- have negative $a_s(0)$~\cite{bartschat2002ultralow}, and it is reasonable to assume that Yb will behave similarly.
This is consistent with the sole calculation of the low energy Yb$-e^-$ elastic scattering cross section, which showed a Ramsauer-Townsend minimum at around \SI{80}{\milli\eV}~\cite{dzuba1994correlation}.

The existence of a bound anionic state in the $p$ configuration has been a matter of considerable discussion. 
Although several calculations~\cite{gribakina1992structure,vosko1991predictions,felfli2009resonances,felfli2020low} and early experimental evidence~\cite{litherland1991observation} suggested that the $p_{1/2}$ state of Yb$^-$ is bound by approximately $\SI{50}{\milli\eV}$~\cite{dzuba1994correlation}, subsequent evaluation~\cite{andersen1999binding,dzuba1998low,o2008valence} points towards Yb$^-$ having a negative electron affinity in both fine structure components. In particular, an experiment at the Aarhus storage ring ruled out an electron affinity greater than $\SI{3}{\milli\eV}$~\cite{andersen1999binding,andersen1998search}, much smaller than the predicted fine structure interval of approximately $\SI{60}{\milli\eV}$~\cite{dzuba1994correlation}. This is consistent with corroborating calculations that Yb possesses shape resonances at $\SI{20}{\milli\eV}$ and $\SI{80}{\milli\eV}$~\cite{dzuba1998low} and complementary evidence obtained from a Rydberg state quenching experiment~\cite{Reicherts2000}.
Despite these efforts, there is still no conclusive evidence for a negative electron affinity in Yb or a measurement of an electron-Yb shape resonance.

\subsection{Rydberg ytterbium Yb$^*$}\label{sec:rydbergytterbium}
We describe the Rydberg atom $\mathrm{Yb}^*$ using a single-channel $LS$ coupling scheme in which the full electronic wave function is written as
\begin{equation}\label{eq:rydbergwft}
\begin{aligned}
   \psi_{\nu_\alpha\alpha}(\vec r,\vec r_c) &= \phi_{\nu_\alpha l_r}(r)\phi_{n_cl_c}(r_c)\\&\times\bra{\hat r,\hat r_c}[(l_rl_c)L_R(s_rs_c)S_R]J_RM_{J_R}\rangle.
\end{aligned}
\end{equation}
Note that $\alpha$ plays the role of a collection of quantum numbers, i.e.\ $\alpha = \{[(l_rl_c)L_R(s_rs_c)S_R]J_RM_{J_R}\}$, where $J_R$ is the total angular momentum of the Rydberg state with $M_{J_R}$ its projection onto the internuclear axis. 
We use $r$ and $c$ to label the Rydberg and core electrons: $\phi_{\nu_\alpha l_r}(r)$ is the radial wave function of the Rydberg electron with orbital angular momentum $l_r$, $\phi_{n_cl_c}(r_c)$ denotes the radial wave function of the core electron, and $s_r=s_c=1/2$ are the corresponding electron spins.
Quantities possessed by the Rydberg atom use an upper case $R$ subscript, i.e., the summed orbital and spin angular momenta $L_R$ and $S_R$.

For the electronic $^1S_0$ states we focus on here, this approximate single-channel description in the $LS$-coupling scheme turns out to be highly accurate, as these levels have a nearly energy-independent quantum defect ($\mu_{^1S_0}\approx\num{4.27}$) and only weak channel mixing ~\cite{peper2025spectroscopy}. 
In contrast, many of the other $^{174}$Yb Rydberg states require a multichannel treatment due to strong spin–orbit coupling and the influence of low-lying doubly excited states.
Recently, a comprehensive multichannel quantum defect theory (MQDT) model for $^{174}$Yb was obtained from a systematic fit to the wealth~\cite{aymar_highly_1980, Ali_1999,lehec2018laser} of spectroscopic data collated in Refs.~\cite{peper2025spectroscopy} and~\cite{peper2020precision}.
This dataset includes Rydberg levels up to and including $L_r = 4$ and spectra for both $^{171}$Yb and $^{174}$Yb. 

This MQDT model quantitatively predicts the breakdown of LS coupling, revealing that the $^1P_1$ states contain between~3 and~6\% $^3P_1$ character admixture, while the $^1D_2$ and $^3D_2$ series are mixed to an even greater degree of about 15\%~\cite{peper2025spectroscopy}. 
Although these Rydberg levels are mixed into the predominantly $^1S_0$ molecular state by $\hat V_a$, the admixture is only at the level of a few percent or less and thus the error in the potential energy curves accumulated by taking this $LS$-coupled single-channel approach will be on the sub-percent level. 

Another significant manifestation of channel interactions in Rydberg spectroscopy is the strong energy-dependence of the quantum defects $\mu_\alpha$ caused by mixing between Rydberg states and low-lying doubly-excited states.
In Yb, this is, for example, seen in the mixing of $f^{13}5d6snd$ $^3P_1$ levels into the $f^{14}6snp$ $^3P_1$ Rydberg series; this causes the $^3P_1$ quantum defects to change by almost $\num{0.4}$ over the range of $n$ explored here. 
Similar changes are seen in the $^3P_0$ series. 
Although we include the energy-dependent quantum defects into our calculation, we neglect the contributions to the Rydberg wave function from these low-lying doubly excited states, which again only enter at the sub-percent level.

\subsection{Methodology}
\label{sec:methodology}
Two different approaches (see Refs.~\cite{giannakeas_generalized_2020,tarana2016adiabatic}  for alternative methods) are typically used to compute the potential energy curves (PECs) of ultralong-range Rydberg molecules. 
The first approach is to diagonalize the molecular Hamiltonian given by Eq.~\ref{eq:hamiltonian} in a basis of electronic Rydberg states~\cite{anderson2014photoassociation,eilesHamiltonian2017}. 
This method is straightforward to implement, can be extended to include the effect of external fields, and gives immediate access to quantities like permanent and transition electric dipole moments; however, the resulting PECs formally do not converge with respect to basis size~\cite{feyComparative2015,eilesHamiltonian2017}, nor is the electronic kinetic energy treated self-consistently in the evaluation of scattering phase shifts~\cite{eiles2023,eiles_trilobites_2019}.
The basis size is thus normally treated as a fitting  parameter when comparing with experimental results~\cite{engelPrecision2019,peper_heteronuclear_2021}, and  the scattering phase shifts used to obtain accurate vibrational spectra are usually basis dependent.%
The extension of this method to a two-electron, single channel Rydberg atom in a strict $LS$-coupling scheme was performed in Ref.~\cite{wojciechowska2025ultralong}; we summarize the key details in Appendix~\ref{sec:appendixDiag}.

\begin{figure}[t]
  \centering
  \includegraphics[width=\columnwidth]{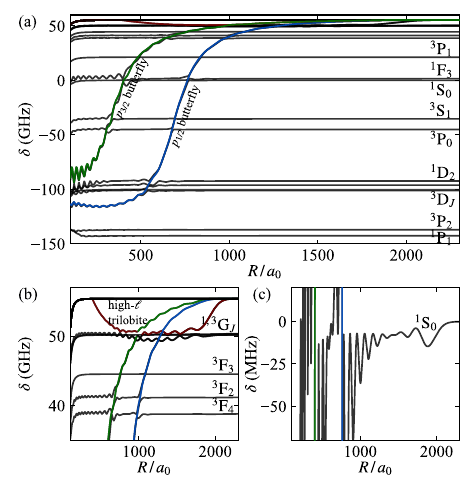}
  \caption{(a)~Overall PEC landscape between the $\nu=31$ and $\nu=32$ degenerate high-$\ell$ manifolds, with the energy axis referenced to the $6s36s\, ^1S_0$ level. 
  The Bohr radius is denoted $a_0$. 
  All dissociation thresholds corresponding to Rydberg $\nu\,^{2S_R+1}{L_R}_J$ states included are shown; on this scale, the corresponding PECs are essentially flat. The dominant features are the twin butterfly PECs (green and blue).
  (b)~Close-up of the region near the $\nu=32$ high-$\ell$ hydrogenic manifold with the nearby $^3F_J$ and $^{1,3}G_J$ levels and the trilobite PEC (red).
  (c)~Close-up of the $6s36s\, ^1S_0$ PEC.}
 \label{fig:overview_all_pecc}
\end{figure}

The second method utilizes the closed-form solution of the Coulomb Green's function to derive a determinantal equation for the Rydberg electron's energy in the field of both the Coulomb potential and the neutral atom~\cite{hamilton2003photoionization,eiles2023}.
We have extended this treatment to Yb within the single-channel approximation, and describe in Appendix~\ref{sec:appendixGF} the relevant modifications.%
This method naturally includes the influence of the whole Rydberg spectrum and thus resolves the convergence issues at the expense of a more inflexible numerical implementation.
We therefore employ this method for all quantitative analyses, using the diagonalization method only to obtain the overview of all PECs shown in Fig.~\ref{fig:overview_all_pecc}(a).

\subsection{Potential energy landscape}\label{sec:potentialenergylandscape}

Figure~\ref{fig:overview_all_pecc} shows PECs computed by diagonalizing $\hat H_{el}$ (Eq.~\ref{eq:hamiltonian} ) in a basis of Rydberg states with $31\le \nu \le 32$ and $M_J=0$, thereby describing the landscape around the $6s36s\, ^1S_0$ state.
The calculation uses the electron–atom scattering phase shifts obtained as a result of our spectroscopic analysis described in Sec.~\ref{sec:results}.  
The PECs fall into two general categories depending on the quantum defect of the unperturbed Rydberg state.  
The first category consists of electronic states with large quantum defects, or equivalently, low angular momentum $(L_R\le 4)$. 
These states are only weakly perturbed, and they appear nearly flat on the scale of Fig.~\ref{fig:overview_all_pecc}(a).  
More dramatically visible are the PECs belonging to the second category, the so-called ``trilobite" and spin-orbit split ``butterfly" PECs. 
These form due to the hybridization of the nearly degenerate high angular momentum Rydberg states by the $s$ and $p_j$-wave interactions, respectively.
Each of these PECs is approximately doubly degenerate, composed of one curve of approximately triplet and one of approximately pure singlet Rydberg character.
The small splitting between these curves, essentially invisible in the trilobite curve (see panel (b)) and largest in the $p_{3/2}$ butterfly curves, stems from the fine structure splitting of the $p$-wave scattering interaction. 
This couples the singlet and triplet Rydberg states, breaking the degeneracy. 
Another notable feature shown in panel~(a) is the strong dependence of the quantum defects not just on the orbital angular momentum, as in alkali atoms, but additionally on $S_R$ and $J_R$. 

Figure~\ref{fig:overview_all_pecc}(b) focuses on the behavior of the trilobite and butterfly curves in the vicinity of the hydrogenic manifold.
Close inspection of the avoided crossings between the butterfly potential curves and the $^{3}F_J$ potential curves reveals again the nearly degenerate doublet of butterfly states, as the triplet-dominated butterfly curves exhibit visible avoided crossings while the singlet-dominated butterfly curve passes through nearly unperturbed.  
The wells in the trilobite curve are significantly disrupted by the energetically adjacent $^3G_J$ potential curve.

Figure~\ref{fig:overview_all_pecc}(c) shows the PEC most relevant to our measurements, namely, that of the $^1S_0$ Rydberg state. 
This curve mirrors, to a first approximation, the electronic probability density, and molecular bound states exist as quantized levels bound in between the nodes of the wave function since $a_s(0)$ is negative. 
The wave function lobe closest to the classical turning point, at approximately $\num{2000}\,a_0$, is typically the deepest, broadest, and least affected by the energy-dependence of the scattering length and higher-order partial waves. 
As a result, the ground vibrational state in this outermost well -- labeled $A0$ in the following -- tends to have the most universal description.
Its binding energy decreases as $\nu^{-6}$ due to the connection to the normalization of the Rydberg wave function~\cite{anderson2014photoassociation,eiles_trilobites_2019}.

Moving inward, the PEC features additional, shallower wells as the electron energy increases towards the energy of the $p_{1/2}$ shape resonance, causing the $p_{1/2}$-wave interaction to compete with the $s$-wave contribution. 
Closer to resonance, this produces the deep potential wells capable of supporting vibrational states whose binding energies are much larger than that of $A0$. 
At resonance ($R\sim 900a_0$) the PEC drops rapidly. 
This converts most vibrational levels -- usually all but the A0 state in the outermost well -- from bound states into resonances whose lifetimes are stabilized by quantum reflection~\cite{rost2010} and potentially sensitive to non-adiabatic effects~\cite{durst_nonadiabatic_2025}. 
The $p_{3/2}$-wave interaction causes a similar drop at even smaller $R$, creating a parallel set of deep inner wells. 
The calculation of vibrational levels in these PECs is described in Appendix~\ref{sec:appendixBS}.

There is an intimate connection between the energies and wave functions of ULRM bound states and the low-energy scattering phase shifts underlying the PECs.
The dearth of knowledge of the phase shifts from ab-initio calculations can be resolved empirically by ULRM spectroscopy, as we demonstrate in the subsequent sections. 

\section{Measurement of ytterbium ULRM spectra}
\label{sec:experiment}

\begin{figure*}
  \centering
  \includegraphics[width=\textwidth]{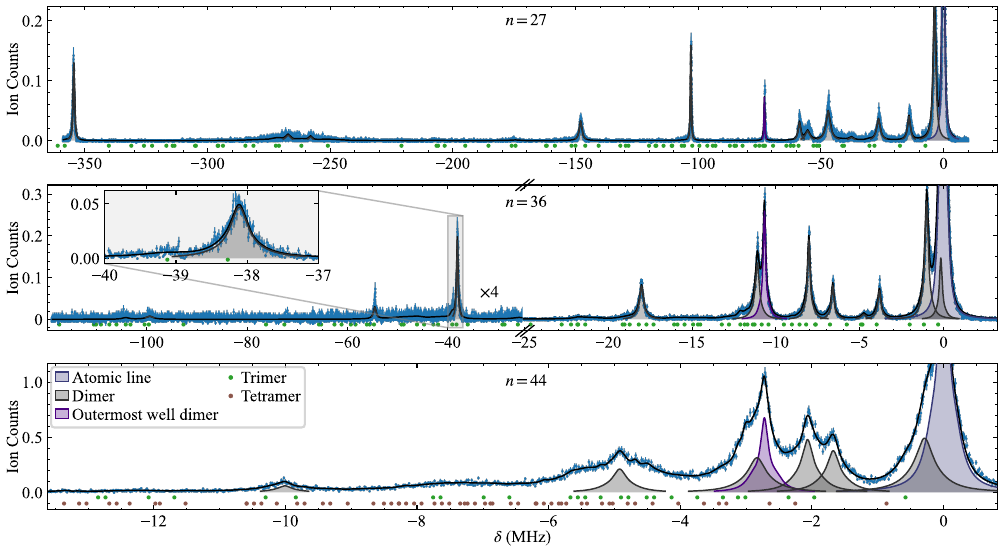}
  \caption{Spectra near the atomic $6sns\,^1 S_0$ state for $n=27$, $n=36$ and $n=44$. 
  Blue points show the mean ion counts versus detuning~$\delta$ from the atomic Rydberg line at $\delta=0$; error bars indicate statistical uncertainties.
  A multi-Lorentzian fit (black) models the atomic transition, the dimer lines, and polyatomic molecular lines (trimers and tetramers) at binding energies given by sums of integer multiples of the corresponding dimer binding energies.%
  Atomic and dimer contributions to the fit are highlighted as shaded curves, and the polyatomic binding energies are marked by points below the data. The vertical axis for $n=36$ is scaled by a factor of \num{4} for $\delta<\SI{-25}{\mega\hertz}$ to enhance visibility.}
  \label{fig:example_spectrum_fits}
\end{figure*}

To produce and detect ULRMs of ytterbium in our experiment, we prepare a dense ultracold cloud of $^{174}\mathrm{Yb}$ atoms in a \qty{1070}{\nano\meter} crossed-beam optical trap. 
In a first step, a two-color magneto-optical trap (MOT) cools $\sim\! 10^7$ atoms below \qty{10}{\micro\kelvin}, as described in Ref.~\cite{urunuela2025}. 
With a subsequent step of evaporative cooling and compression in the optical trap, we reach temperatures of \qty{4}{\micro\kelvin} at atomic densities up to $\rho=\qty{4e14}{\per\cm\cubed}$.
These conditions ensure that the thermal energy remains below the small binding energies of ULRMs, which can be as low as $\sim\!\SI{200}{\kilo\hertz}$, and that there is a substantial probability of finding two atoms at separations comparable to the characteristic radius of the Rydberg electronic wave function.
For the range of studied Rydberg states, this density corresponds to a mean interatomic distance of $\SIrange[]{0.8}{2.1}{}\, r_\mathrm{Ryd}$, where $r_\mathrm{Ryd}$ denotes the distance from the nucleus to the outer lobe of the Rydberg electronic wave function.
The $^{174}$Yb atom-atom scattering length \SI[separate-uncertainty=false]{5.55(8)}{\nano\meter}~\cite{Julienne2008} is much smaller than the mean interatomic distance.
Thus, two ground state atoms inside the electron orbit do not interact with each other.

As shown in Fig.~\ref{fig:expscheme}(b–c), we excite a $6sns\,^1 S_0$ Rydberg state using two counter-propagating lasers at \qty{399}{\nano\meter} and \SIrange[range-phrase = –]{395}{397}{\nano\meter}, detuned by~$\delta$ from the Rydberg state and by $\Delta \approx \SI{1}{\giga\hertz}$ from the intermediate $6s6p\, ^1P_1$ state.
The excitation is applied in short pulses, with the optical trap switched off during each pulse, and the atoms are recaptured afterward.
Up to \num{8000} pulses are applied to the atomic cloud prepared in each experimental cycle.
By varying the detuning~$\delta$ from pulse to pulse, we probe the binding energies at which ULRMs photoassociate.
We use Tukey-shaped pulses of \qty{32}{\micro\second} (FWHM) duration.
To minimize power broadening and the probability of creating multiple Rydberg excitations, we use two-photon single-atom Rabi frequencies in the range of \SIrange{350}{800}{\hertz}.
This results in a maximal spectral resolution of $\sim\!\qty{37}{\kilo\hertz}$.

Eight in-vacuum electrodes, arranged as ring segments above and below the atomic cloud (Fig.~\ref{fig:expscheme}(c)), provide shielding from stray electric fields.
They also allow active compensation, yielding zero-field conditions to within $\SI{1}{\milli\volt\per\centi\meter}$.
After each spectroscopic pulse we detect excitations of Rydberg atoms or molecules via electric-field ionization.
To this end, we apply a field of $\sim\!\qty{70.6}{\volt\per\cm}$ using two electrodes positioned above the atomic cloud.
An additional electrode guides the resulting ions to a microchannel plate (MCP) detector, providing a low-background spectroscopic signature.
We note that for states $n<47$ the applied ionization field lies below the field ionization threshold, but we still detect a considerable signal down to $n=26$, as further discussed in Appendix~\ref{appendix:A}. 

By averaging several hundred scans, we obtain spectra with a high signal-to-noise ratio.
Figure~\ref{fig:example_spectrum_fits} presents three examples for representative Rydberg states.
Using the same procedure, we measured Rydberg-molecule spectra for $n=\numrange[range-phrase = -]{26}{27}$ and $\numrange[range-phrase = -]{30}{45}$.
The spectra contain the atomic Rydberg line along with lines at lower energies corresponding to diatomic Rydberg molecules (dimers).
The atomic line position is determined from separate low-density scans, where molecular formation is suppressed, and defines the zero of the detuning~$\delta$ axis (see Appendix~\ref{appendix:B} for details).
In fact, the probability of finding a ground state atom within the Rydberg volume scales as $\rho r_\mathrm{Ryd}^3\propto\rho\nu^6$~\cite{pfau2014}. 
In an uncorrelated gas, the probability of finding $N$ atoms within the binding region scales as the $N$-th power of the single-atom probability, so that at higher densities or for higher $n$, polyatomic molecules composed of one Rydberg atom and two or more ground-state atoms contribute additional lines to the spectrum~\cite{rost2010,pfau2014}.
The binding energies of the polyatomic molecules are given by integer linear combinations of the binding energies of the constituent diatomic molecular states, reflecting the additive molecular potential of $S$-state Rydberg molecules~\cite{rost2010,pfau2014,camargo_creation_2018}.

We fit the spectra using a sum of spectral lines that account for deviations from a purely Lorentzian line shape.
Polyatomic peaks are modeled with binding energies constrained to sums of the corresponding dimer energies.
For states with $n<42$, we include up to triatomic molecules (trimers), and for $n\ge 42$, up to tetramers.
This fitting procedure requires prior knowledge of which lines correspond to dimers.
We distinguish dimer lines from polyatomic contributions by their different scaling of peak amplitude with atomic density.
Details of the fitting procedure and dimer classification method are provided in Appendix~\ref{appendix:B}.
As illustrated by the example spectra in Fig.~\ref{fig:example_spectrum_fits}, polyatomic contributions are clearly visible at $n=44$, while being much weaker at $n=36$, and nearly absent at $n=27$.
At relatively high Rydberg states (e.g., $n=44$), the spectral signatures of polyatomic molecules are so densely spaced that they cannot be resolved individually and also obscure any weak dimer lines that might be present.

Moreover, with increasing $n$, the molecular binding energies decrease rapidly and the spectral lines become more closely spaced, resulting in partial overlap at higher $n$.
Specifically, the dimer bound in the outermost potential energy well ($A0$) can be unambiguously identified through its characteristic $\nu^{-6}$ binding-energy scaling (Sec.~\ref{sec:potentialenergylandscape}).
Additional dimer lines at larger binding energies provide clear experimental signatures of the $p$-wave shape resonance.

For scans with the largest number of excitation pulses, we observe a progressive density reduction of up to \SI{60}{\percent} across the pulse train, caused by absorption and heating on the resonance of the two-photon transition.
This density reduction does not affect the extracted molecular binding energies.

\section{Results}\label{sec:results}
In this section, we describe how the electron-Yb scattering phase shifts are extracted from the molecular spectroscopy detailed in the previous section, how the rich vibrational structure is analyzed and categorized with help from the theoretical model, and how this can be used to further fine-tune the phase shifts. 
Finally, we show how detailed information about the Rydberg atom's structure is revealed in the vibrational spectra. 
\subsection{Fitting of scattering phase shifts}
Theoretical calculation of the vibrational spectra using the methods discussed in Sec.~\ref{sec:methodology} requires knowledge of the electron-Yb scattering phase shifts. 
As neither these nor the electron-Yb interaction potential are known, we describe the latter with the $L_a$-dependent model potential~\cite{bahrim2001boundary,bahrim2002near,peper2020precision}
\begin{equation} 
\label{eq:model_potential}
V_{L_a}(r)=-\frac{L_aZ_c}{r}e^{-\lambda r}-\frac{A e^{-\gamma r}}{r^{1-L_a}}-\frac{\alpha}{2r^4}\left(1-e^{-(r/r_c)^6}\right). 
\end{equation} 
The first two terms of Eq.~\ref{eq:model_potential} account for the short-range Coulomb interaction, screened by the core electrons, while the third term describes the long-range polarization potential between the electron and the atom, multiplied by a cutoff function to avoid the divergence at the origin. 
This uses in total eight fit parameters: the polarizability $\alpha$, the $L_a$-dependent parameters  $A$, $\gamma$, and $r_c$, and $\lambda$ for $L_a=1$. 
$Z_c$ is the nuclear charge. 
We include the standard spin-orbit interaction and numerically integrate the radial Schrödinger equation with the potential 
\begin{equation}
\label{eq:so}
   V(r) = V_{L_a}(r)+\frac{1}{2c^2r}\frac{\mathrm{d}V_{L_a}(r)}{\mathrm{d}r}\vec l_r \cdot \vec s_r
\end{equation}
to obtain the scattering phase shifts for the $s_{1/2}$, $p_{1/2}$, and $p_{3/2}$ symmetries.
Here, $\vec l_r$ and $\vec s_r$ are the orbital and spin angular momentum of the Rydberg electron~\cite{bahrim2002near} in the frame of the perturber atom~\footnote{These labels should not to be confused with $l_r$ and $s_r$ from Eq.~\ref{eq:rydbergwft}}. 
Some details relating to the treatment of the divergence of Eq.~\ref{eq:so} at short distances are provided in Appendix~\ref{sec:appendixPS}, where we closely follow the approach of Ref.~\cite{bahrim2001boundary}.

We employ the Green's function method to compute the PECs, which are subsequently used to determine the vibrational energies and wavefunctions, as described in Appendix~\ref{sec:appendixBS}.
We first obtain the eight fitting parameters for Eq.~\ref{eq:so} by minimizing the root mean square difference between calculated and measured binding energies for those states more weakly bound than the $A0$ state for $n=31$ and $n=36$.
The $n=31$ Rydberg state was chosen for its relatively clean vibrational spectrum, while the $n=36$ spectrum contains a second state nearly degenerate with the $A0$ state. 
This provides a tight constraint on the relative contributions of $s$- and $p$-wave scattering channels.  
As described in Appendix~\ref{sec:appendixERT}, the effective range theory for a polarization potential~\cite{o1962low,o1964low} proves useful for identifying initial estimates for the fit parameters, which we then optimize using a gradient descent algorithm. 
After optimization for these two $n$ levels, we use the resulting phase shifts to calculate vibrational spectra for all measured Rydberg states.
This global comparison over a broad span of $n$ values helps to avoid overfitting or systematic bias in the fit, and increases the range of scattering energies over which the fit parameters can be verified. 

\subsection{Spectroscopic assignment}

\begin{figure*}[t]
  \centering
  \includegraphics[width=\textwidth]{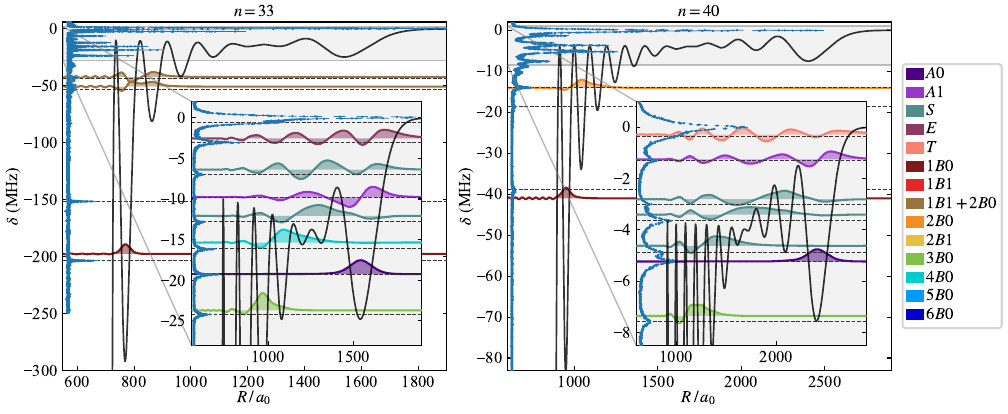}
  \caption{Calculated PEC (black) and vibrational wave functions near the $^1S_0$ asymptote for $n=33$ and $n=40$. The Green's function formalism was used to obtain these PECs. The amplitude of each wave function is shown at its corresponding eigenenergy. The color scheme identifies different classes of molecular states (see main text). The measured spectra (blue) and the fitted experimental energies (gray dashed lines) are shown for comparison.
  The insets provide a zoomed-in view of the region between the atomic line and the outermost-well ground state.}
  \label{fig:exp_theo_comp}
\end{figure*}

\begin{figure*}
  \centering
  \includegraphics[width=\textwidth]{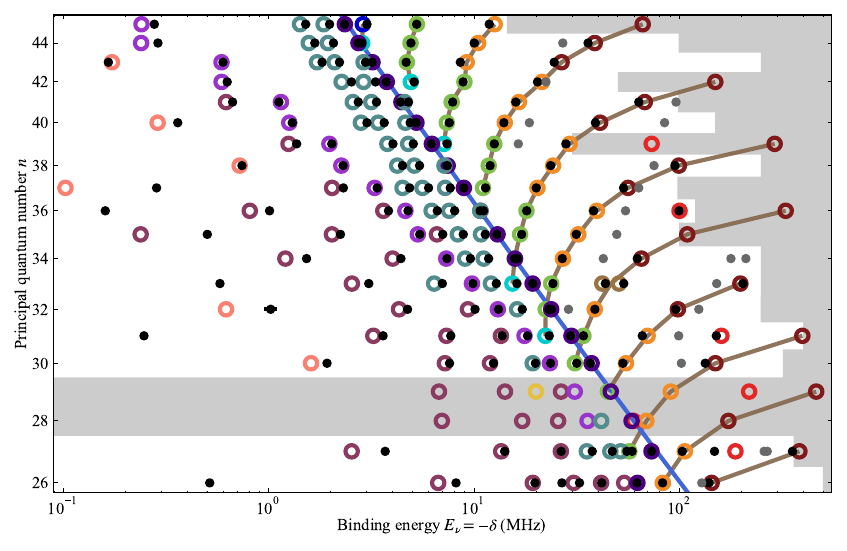}
  \caption{Measured and calculated molecular binding energies for $n=\SIrange[]{26}{45}{}$.
  Measured energies are shown as black points, while calculated energies are indicated by colored circles following the color code defined in Fig.~\ref{fig:exp_theo_comp}.
  The straight blue line shows a power-law fit to the measured binding energies of the ground state in the outermost potential well~$A0$.
  Light-brown curves highlight systematic scalings of the deeply bound states localized in the $p_{1/2}$ ``butterfly" wells.
  Gray points mark lines missing in the model prediction that are likely bound in the $p_{3/2}$ ``butterfly" wells (see main text).
  Error bars on the experimental energies represent the $1\sigma$ confidence intervals of the fit and are mostly smaller than the marker size.
  Experimentally unexplored regions are shaded in gray.
  }
    \label{fig:overview_all_states}
\end{figure*}
Two exemplary comparisons of the theoretical and observed spectra are presented in Fig.~\ref{fig:exp_theo_comp}. 
The experimentally obtained dimer peak positions (Sec.~\ref{sec:experiment}) are shown as dashed lines, and the calculated vibrational states are plotted centered around their binding energies. 
To facilitate a systematic analysis and highlight trends in the spectrum, we categorize the vibrational levels according to the following scheme based on the character of the calculated vibrational states.
The state labels and color codes are introduced in the figure legend. 
The well-localized ground vibrational state in the outermost well, $A0$, has already been introduced in Sec.~\ref{sec:potentialenergylandscape}.
The $p$-wave contribution lowers the inner barrier of this potential well, causing its excited states to delocalize across several wells.
We label the state that most closely resembles the first excited state of the pure $s$-wave potential's outermost well as $A1$. 
Excited states that are bound due to quantum reflection from the rapid potential drop are labeled $E$.
States near the dissociation threshold ($T$) are very sensitive to the phase shifts, as only tiny changes in the PECs to make them shallower will push these states into the continuum.
The examples $n=33$ and $n=40$ in Fig.~\ref{fig:exp_theo_comp} illustrate this sensitivity: threshold states are observed experimentally in both cases but reproduced by the model only for $n=40$.
Another class of states is bound in the ``shoulder" of the outermost well and is labeled $S$. 
It is one of these $S$ states which is nearly degenerate with $A0$ in several Rydberg states including $n=40$. 

Vibrational states localized in the deep $p_{1/2}$-wave-affected wells are labeled $iBj$.
Here, $i$ is the index of the well in which the state is predominantly localized (counted outward), while $j$ denotes the vibrational excitation order in that well.
For example, for $n=40$, the ground state in the first well lying around $\SI{-40}{\mega\hertz}$ is labeled $1B0$, while the ground state in the second well is labeled $2B0$.
As $n$ increases, the difference between these states and shoulder states becomes increasingly ambiguous.
For $n=33$, in addition to the $1B0$ and $3B0$ states, an interesting phenomenon occurs in the first and second deep wells: the first excited state of the first well ($1B1$) and the ground state of the second well ($2B0$) are nearly degenerate and therefore hybridize.
We label the resulting hybrid pair of states as $1B1+2B0$.
Not all observed states -- the one at $\SI{-150}{\mega\hertz}$ for $n=33$, for example -- could be assigned in this fashion as they lack a corresponding state in the calculation.

\subsection{Trends in vibrational binding energies}
To better understand these unassigned levels, we turn to 
the comprehensive comparison of vibrational energies across the investigated range of $n$ summarized in Fig.~\ref{fig:overview_all_states}. 
Throughout, we employ the same color-coded state classification as in Fig.~\ref{fig:exp_theo_comp}, which highlights the distinctive $n$-dependence of the molecular level structure for both $s$- and $p_{1/2}$-wave dominated vibrational states. 
A fit to the experimental ground-state energies in the outermost well ($A0$) for $30\le n\le 45$ yields a scaling exponent of $\num[separate-uncertainty = false]{-5.9453 \pm 0.0004}$, close to the $\nu^{-6}$ scaling expected from Rydberg wave function scalings. 
We return to the exceptional case of $n=26$ and its deviation from this scaling in Sec.~\ref{sec:n26}.
The structure of the more weakly bound states ($A1$, $E$, $T$) remains qualitatively unchanged across different values of $n$, with the binding energies following a similar scaling.
For states with $n \ge 38$, we impose a cutoff on the Franck–Condon factors of the vibrational states to filter out a small number of excited $A1$ or $E$ states with negligible coupling to the initial state. 
These are approximately anti-symmetric with respect to reflection about their center, such as 
the state $A1$ in $n=33$.
While this one is observed experimentally, its line strength is significantly weaker than that of the other states. 
At larger $n$, these states are not observed due to the increasingly dense level spacing and the enhanced background from polyatomic molecules.

Another trend is the overall staggering of binding energies larger than that of the $A0$ state. 
Here, we see the effect of a mismatch in $\nu$ scaling between the $R$-position of the $p_{1/2}$-wave resonance, which is approximately linear in $\nu$ over this range, and the $R$-position of the $p$-wave-dominated wells.
As $n$ increases, the $R$-position of the $p_{1/2}$-wave resonance moves outward faster than the potential wells just outside it.
Consequently, the wells near the approaching shape resonance are increasingly pulled down, causing the binding energies of the states localized in them to rapidly increase, as highlighted by the brown guidelines in Fig.~\ref{fig:overview_all_states}.
The wells nearest the steep drop associated with the shape resonance are pulled down more strongly, producing a more rapid increase in binding energy with $n$ than in wells farther out.
With further increase in $n$, the shape resonance eventually overtakes and removes the innermost well, thereby terminating the series. 
This also makes the next outer well the new innermost well, shifting the state labels $iB0$ in the remaining wells to $(i-1)B0$.
A similar structure, though less complex, was observed in Ref.~\cite{engelPrecision2019}, where the energies of two $p$-wave-dominated states in Rb were found to alternate as a function of $n$.
Interestingly, some states localized in the $p_{1/2}$-wave-dominated wells have binding energies smaller than those of the outermost well ground state, causing the brown trendlines to frequently cross the blue trendline of the $A0$ states.
For the hybridized pair at $n=33$ ($1B1+2B0$), we draw the trendline through their mean binding energy. 
The first excited states in the $p$-wave-affected wells, $iB1$ (red and yellow), appear only when the corresponding $iB0$ states are very deeply bound, causing them to appear too sporadically in $n$ to reveal similar trends.

Only a few experimental levels, marked in gray on the figure, avoid categorization in this way.
These all have binding energies higher than that of the $A0$ state.
We hypothesize that these states are bound in the $p_{3/2}$-wave-influenced potential, as we can confidently rule out binding in the $p_{1/2}$ butterfly wells and because they appear to follow trendlines similar to those of the $iB0$ $p_{1/2}$-well states.
The slower increase in binding energy with increasing $n$ is consistent with these states being localized in $p_{3/2}$-resonance-dominated wells further inward, which move outward more slowly as $n$ increases.

To further support this, we perform a second iteration of the fitting procedure including the deeply bound states in the root mean square error calculation comparing theory and experiment.
In a first pass, we include the missing states in the fit or exchange them for the $1B1$ level (to be concrete: for $n=31$ we attempt to assign the missing state at $\sim \SI{-100}{\mega\hertz}$ to the $1B1$ state, and for $n=36$ we do the same for the missing state at $\sim \SI{-90}{\mega\hertz}$). 
This does not reproduce the agreement across all $n$s and for all excited states satisfactorily. 
We then exclude the unassigned peaks and include all labeled peaks in the RMS error. 
The resulting fit performs very well, and allows us to refine the extracted phase shifts further due to the extreme sensitivity of these deeply bound levels to the position and width of the $p_{1/2}$ shape resonance. 

\subsection{Extracted scattering phase shifts}
\begin{figure}[t]
  \centering
  \includegraphics[width=\columnwidth]{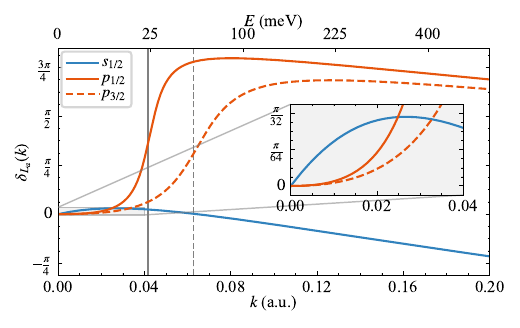}
  \caption{Energy-dependent phase shifts for the $s_{1/2}$, $p_{1/2}$ and $p_{3/2}$ partial waves. The inset shows the low-energy region.
  Vertical gray solid (dashed) lines denote the positions of the $p_{1/2}$ ($p_{3/2}$) shape-resonances.}
  \label{fig:phase_shifts}
\end{figure}

With the phase shifts calculated from this second fit, the agreement between experiment and theory shown in Fig.~\ref{fig:overview_all_states} is excellent. 
The only experimentally observed states completely missed by the model are the very sensitive threshold states for $n=31$, 33, and 36. All other levels agree to within 10\% relative error in the binding energy. 
From this we conclude a high level of confidence in the fit parameters of the model potential and the resulting energy-dependent phase shifts shown in  
Figure~\ref{fig:phase_shifts}. 
The two $p$-wave shape resonances are apparent:  the $p_{1/2}$ resonance lies at $\SI{23.7}{\milli\eV}$, consistent with the value of $\SI{20\pm10}{\milli\eV}$ predicted in Ref.~\cite{dzuba1998low}, and the broader $p_{3/2}$ resonance lies at $\SI{53.5}{\milli\eV}$. 
This fine-structure splitting of $\SI{30}{\milli\eV}$ is approximately half the separation reported previously~\cite{dzuba1994correlation}. 
We estimate the uncertainties in the resonance positions to be approximately \SI{0.5}{\milli\eV} for the $p_{1/2}$ resonance and \SI{20}{\milli\eV} for the $p_{3/2}$ resonance. These estimates are obtained by initializing the fitting procedure from a range of starting parameters to explore alternative local minima in the optimization landscape. Among the solutions that yield comparably good agreement between experiment and theory for the butterfly states, we observe a corresponding spread in the extracted $p$-wave resonance positions, which we take as a measure of the uncertainty.

The zero-energy $s$-wave scattering length is determined to be $a_s(0)=\num{-7.5}\,a_0$. 
To estimate the uncertainty in $a_s(0)$, we take the first-order perturbation theory result for the outermost potential well, $U_n(r) \approx 2\pi a_s(0)|\psi_{\nu 00}(R)|^2$, and assume that the ground-state energy in this potential well is proportional to the well depth over this range of $n$.
From this, we convert the residuals from the difference between experimental and theoretical binding energies into an uncertainty $\Delta a_s(0)$ in the scattering length for each $n$, and compute the mean value of this to obtain $\Delta a_s(0)\sim 0.02\,a_0$.
To account for the several assumptions made in this estimation, particularly the neglect of $p$-wave contributions, we take a more conservative value of $\pm 0.1\,a_0$.
The fitted value for the polarizability, $\alpha=\num[separate-uncertainty=false]{141.2\pm0.05}$, agrees with the recommended value computed in Ref.~\cite{dzuba2010dynamic}, although our estimated uncertainty is an order of magnitude smaller. 
The Ramsauer-Townsend zero predicted from our fit results lies at $\SI{56}{\milli\eV}$. 
As our vibrational assignment requires the existence of a shape resonance in both $p_j$ channels, we conclude that Yb$^-$ does not exist. 

The molecular states investigated are most sensitive to details of the $s_{1/2}$ and $p_{1/2}$ scattering phase shifts, and thus firm conclusions can be drawn only for these channels.
Accordingly, different parameter sets may reproduce nearly the same $s_{1/2}$ and $p_{1/2}$ phase shifts while differing in the $p_{3/2}$ channel.
To further constrain the $p_{3/2}$-channel parameters, the ULRMs localized in the inner butterfly potential could be investigated. 

\subsection{Refinement of $^1F_3$ quantum defect}\label{sec:n26}

\begin{figure}[t]
  \centering
  \includegraphics[width=\columnwidth]{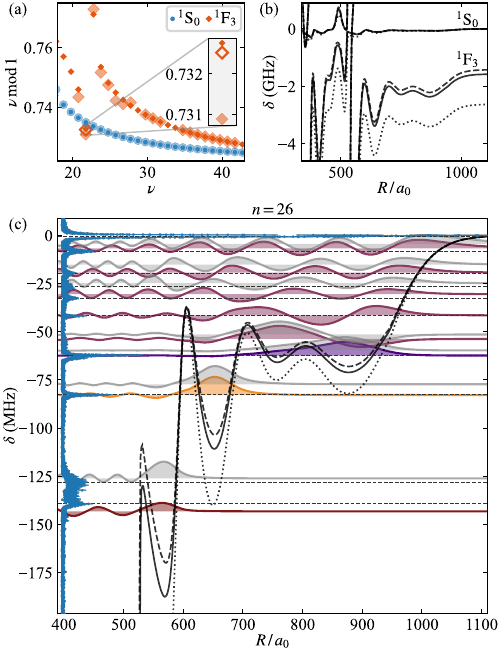}
  \caption{(a)~Lu-Fano-type plot showing the fractional part of the effective quantum number~$\nu$ over the range of interest for the $^1S_0$ and $^1F_3$ Rydberg series.
  Solid markers represent the best current values from the MQDT fit in Ref.~\cite{kuroda2025microwave}, extracted using \emph{pairinteraction}~\cite{Hofferberth2017}, based on experimental values~\cite{robaux1984,lehec2018laser,kuroda2025microwave} shown in lighter color.
  The open diamond corresponds to the quantum defect obtained in this work.
  (b)~PECs near the $6s26s\,^1S_0$ asymptote, showing the nearby $^1F_3$ curve using the $^1F_3$ quantum defect extracted from this work (solid line), the experimental value reported in Ref.~\cite{robaux1984} (dotted line), and the value obtained from the MQDT fit of Ref.~\cite{kuroda2025microwave} (dashed line).
  The $^1F_3$ state potential is substantially deeper than that of the $^1S_0$ state due to the fact that its Rydberg wave function is more localized along a fixed direction, enhancing the binding. 
  (c)~Close-up view of the $^1S_0$ PEC for $n=26$.
  Vibrational states bound by the potential curve which provides the best fit to the experimental data are plotted in the same manner as in Fig.~\ref{fig:exp_theo_comp}.
  Vibrational states in light gray represent those bound in the PEC with the quantum defect from Ref.~\cite{kuroda2025microwave}.
  }
  \label{fig:pec26}
\end{figure}

The molecular spectrum at $n=26$ deviates noticeably from the identified trendlines: in particular, no experimental line appears at the position of the $A0$ state expected from the $\nu$-scaling.
This arises from the strong energy dependence of the $^1F_3$ quantum defect near the $6s26s\,^1S_0$ state, where the $^1F_3$ energy jumps from a value above the $^1S_0$ state to one just slightly below it.
The effect is illustrated in Fig.~\ref{fig:pec26}~(a) in terms of the non-integer part of the effective principal quantum number $\nu \bmod 1$ as a function of $\nu$. 
For $n=26$, this means that the relevant $^1S_0$ and $^1F_3$ potential curves are separated by less than $\SI{3}{\giga\hertz}$ and, as both are singlet Rydberg states, may experience significant mixing due to the $s$-wave interaction with the ground-state perturber. 

In this range of principal quantum numbers, the most accurate experimental $^1S_0$ quantum defects are reported in Ref.~\cite{lehec2018laser}.
More recently, Ref.~\cite{kuroda2025microwave} measured $^1F_3$ energies with high precision using microwave spectroscopy, but only down to $6s29f\,^1F_3$. 
For lower-lying states, the best available data, which are significantly less precise, come from Ref.~\cite{robaux1984}; we use these after applying a global $\SI{-2.4}{\giga\hertz}$ correction, as discussed in Ref.~\cite{kuroda2025microwave}.
However, using the experimental quantum defects produces a theoretical molecular spectrum that disagrees with the experimental data, even for $A0$, to a far larger extent than for any of the other $n$ values investigated.
Furthermore, the calculated binding energies are systematically deeper than the measured values.

Subsequently, we employ the MQDT-fit defects from Ref.~\cite{kuroda2025microwave} (darker symbols in panel (a)) for both the $^1S_0$ and $^3F_1$ energy levels.
This produces significantly closer, but still unsatisfactory, agreement with the experimental data.
To achieve the same level of agreement for the $6s26s\,^1S_0$ state as for other Rydberg states, we then adjust the corresponding $6s23f\,^1F_3$ quantum defect, using the experimental defect for $^1S_0$.
The optimal value is determined by fitting the vibrational state energies to the experimental spectra.

Figure~\ref{fig:pec26}(b) and~(c) compares the resulting PECs for the three cases: the best-known experimental quantum defects (dotted), the MQDT-fitted defects from Ref.~\cite{kuroda2025microwave} (dashed), and our refined $^1F_3$ quantum defect (solid).
As shown in panel~(b), these three quantum-defect values modify the corresponding $^1F_3$ PEC on the scale of~\si{\giga\hertz}, which in turn stretches the $^1S_0$ PEC in panel~(c) on the order of~\SI{10}{\mega\hertz}.
Only for the refined $^1F_3$ defect do the calculated vibrational levels coincide with the experimentally observed lines.%
The optimal value corresponds to $\nu_{^1F_3} = \num[separate-uncertainty=false]{21.73253 \pm 0.00004}$, whereas even the nearby MQDT-fit value $\nu_{^1F_3} = \num{21.73277}$ produces vibrational states (shown in light gray) that are clearly inconsistent with the experimental data.
We note that the extracted quantum defect may carry a systematic bias stemming from the single-channel description of the Rydberg atomic wavefunctions, and its absolute accuracy remains to be validated by a full MQDT treatment. Nonetheless, the pronounced sensitivity of the vibrational spectra to the $6s23f\,^1F_3$ defect constrains its value with high precision.
Furthermore, our analysis is primarily sensitive to the relative $^1F_3$-$^1S_0$ quantum defect, so the extracted $^1F_3$ value depends on the better-known $^1S_0$ energy.
We again observe a few additional vibrational states not accounted for by the $s$- and $p_{1/2}$-wave-dominated PEC, suggesting binding in the $p_{3/2}$-wave-dominated PEC.
These results demonstrate that, once the interaction with the ground-state perturber is accurately modeled, ULRM spectroscopy can provide a sensitive and high-precision probe of Rydberg series that are experimentally challenging to access due to dipole-selection rules.

\section{Conclusion}
In conclusion, we have presented a comprehensive survey of the vibrational spectrum of ultralong-range Yb Rydberg molecules from $n=\SIrange{26}{45}{}$. 
This study is the first of its kind performed in Yb, and illustrates how ULRMs -- beyond the inherent interest in their exaggerated properties -- serve as a high-precision probe of atomic properties. 
After collating vibrational spectra from almost twenty Rydberg levels, we could fit the hitherto unknown electron-Yb scattering phase shifts to a high level of accuracy, obtain the zero-energy $s$-wave scattering length to a few percent uncertainty, and constrain the atomic polarizability to an order of magnitude smaller uncertainty than previous estimates. 
Yb boasts the smallest $s$-wave scattering length, at only $a_s(0)=\num{-7.5}\,a_0$, of any species used to form ULRMs to date. 
Moreover, the observed pattern in the deeply bound vibrational levels gives a clear indication of a $p_{1/2}$ scattering shape resonance, whose position we could identify on a $\si{\milli\eV}$ level of accuracy. 
The existence of additional vibrational states at large binding energies, which cannot be matched to any levels in the $s$- and $p_{1/2}$-wave-influenced potential curves, provides a similarly clear signature of the $p_{3/2}$ shape resonance. 
Although we did not attempt to fit these states directly and thus do not impose tighter constraints on the resonance position, we see that the $p_{3/2}$ scattering channel must also have a resonance lying, with significant uncertainty, a few tens of $\si{\milli\eV}$ away from the $p_{1/2}$ resonance position. 
Finally, our analysis of the exceptional case of $n=26$ revealed how ULRM spectroscopy can be used as a precision measure of relative energy differences between accidentally near-degenerate levels. 

By focusing on the $^1S_0$ electronic states, we have been able to accurately study the ULRMs using a purely single-channel quantum defect theory approach within a strict $LS$ coupling scheme. 
This was especially important in order to obtain accurate information about the $s$- and $p_j$-wave scattering phase shifts without obfuscation.
Future work should extend this framework to a full multichannel treatment of Rydberg molecules, enabling exploration of the much richer structure of multichannel Rydberg vibrational spectra.
Such an approach will not only allow ULRM spectroscopy to refine MQDT fits by taking advantage of the vibrational states’ sensitivity to the detailed composition of the electronic wave function, but will also lead the way towards creating more exotic molecular states using the abundance of near degeneracies and strong channel mixing in Yb~\cite{eiles2015ultracold}.
It would be particularly interesting to explore how the split hyperfine thresholds in isotopes such as $^{171}\mathrm{Yb}$ influence molecular structure. 
With tweezer arrays offering control over the molecular geometry~\cite{guttridge_individual_2025}, they could be used to study ULRMs of divalent atoms at the single-molecule level. 
Finally, by again taking advantage of the near-degeneracy between the $^1F_3$ and $^1S_0$ levels highlighted in this work, it should be possible to create $^1F_3$ molecules even using a two-photon excitation scheme, thanks to the small $S$-state admixture.
This would also provide insight into the singlet-triplet mixing with the $^3F_3$ state, and the resulting coupling of electronic states with different parity could be fruitfully explored under an external electric field.

The evidence we have found for polyatomic Rydberg molecules bound in the additive Rydberg $S$-state potential should be further explored, as the small $a_s(0)$ revealed by our analysis results in  the crossover from few- to many-body physics occuring at lower principal quantum numbers than in other species \cite{pfau2014}. 
As $^{174}\mathrm{Yb}$ lacks hyperfine structure but possesses a $p$-wave shape resonance, it provides an interesting counterpoint to $\mathrm{Rb}$ (which has both) and $^{84}\mathrm{Sr}$ (which has neither), and could be used to resolve questions about Rydberg loss rates in dense gases~\cite{kanungo2020loss,schlagmuller2016probing}.  
Finally, we expect the promising aspects of multivalent Rydberg atoms to benefit the broader class of long-range molecules, which include Rydberg-ion molecules~\cite{duspayev_long_2020,deiss_long_2021,Pfau2022}, Rydberg-molecule molecules~\cite{rittenhouse_ultracold_2010, gonzalez2015rotational, guttridge_observation_2023}, Rydberg bimolecules~\cite{Sadeghpour2021}, and Rydberg macrodimers~\cite{hollerith_quantum_2019, sasmannshausen_long-range_2016, boisseau_macrodimers_2002}, whose study has largely been restricted to alkali metal atoms.

\begin{acknowledgments}
We thank Frederic Hummel for valuable discussions and Anthea Nitsch for proofreading the manuscript.
This work was supported by the European Union's Horizon 2020 program under the ERC grant SUPERWAVE (grant No.101071882) and by the Deutsche Forschungsgemeinschaft (DFG) within the collaborative research center SFB/TR185 OSCAR, project A8 (No. 277625399). 

\section*{Data availability} 
The data presented in this work are available in the Zenodo repository~\cite{legrand_2025_zenodo}.
The Green's function code used to obtain the PECs shown in this work can be obtained upon reasonable request from \url{meiles@purdue.edu}.
\end{acknowledgments}

\appendix
\section{Sub-threshold ionization signal of $^{174}\mathrm{Yb}$}\label{appendix:A}

Our ion-optics setup for Rydberg detection limits the maximal ionization field applied at the atomic-cloud position to $\sim\qty{70}{\volt\per\cm}$.
This field, applied immediately after the spectroscopic pulse with a rise time of \qty{27}{\nano\second}, is sufficient to reach the classical ionization threshold of the $^1S_0$ series for states with $n \ge 47$. Nevertheless, we observe significant ionization rates of Rydberg atoms and molecules in electric fields substantially below the classical threshold, allowing us to measure molecular spectra down to $n=26$, as shown in Sec.~\ref{sec:experiment}. 

Analysis of the pulsed detection data reveals that this sub-threshold signal does not originate from single pulsed field ionization of the $^1S_0$ states. Instead, we observe that it builds up over successive spectroscopic pulses in our pulse-train sequence. We tentatively attribute this to a combined effect of the diabatic electric-field ramps, the presence of multiple laser fields between pulses, and state-changing atomic collisions, enhancing ionization at low electric fields. For the purposes of this work, we have verified that this mechanism does not shift the molecular peak positions relative to the main atomic line; thus, the extracted binding energies remain unaffected. In particular, we observe no Rydberg blockade or level shifts due to already present Rydberg atoms or previously produced ions, which are removed from the atomic cloud by each field pulse.

In this sub-threshold regime, the ionization rate reduces with decreasing principal quantum number $n$. Consequently, for states with $n\le 35$, we increase the power of the spectroscopic pulse to maintain a reasonable signal-to-noise ratio, at the cost of some spectral broadening. This does not adversely affect our analysis, as the molecular level spacing increases substantially at lower $n$.

\section{Fitting procedure of measured Rydberg molecule spectra}\label{appendix:B}
When extracting the molecule binding energy, we calibrate the frequency detuning $\delta$ axis to the measured position of the atomic Rydberg line. To avoid distortions from molecular contributions, this calibration is performed at atomic densities of $\sim\!\SI{3.5e13}{\per\centi\meter\cubed}$, where no molecular signals are observed. To extract the molecular binding energies, we fit each experimental spectrum with a sum of spectral lines.
The line shape is modeled as a Lorentzian with additional symmetric sidebands of fixed relative amplitude and width, which accurately reproduces the measured spectral lines. These sidebands arise from the frequency-stabilization scheme of the \SI{399}{\nano\meter} excitation laser. They are suppressed by more than a factor of five relative to the carrier and are offset by \SIrange[range-phrase=--]{100}{300}{\kilo\hertz}, thereby having a negligible impact on the spectral resolution.

All dimer peaks are fitted using independent lines, whereas the polyatomic lines have binding energies constrained to the sum of two or more dimer energies. Initial dimer candidates are identified by the density scaling of their amplitudes, extracted from spectra recorded at different atomic densities. After the initial fitting cycle, we assess the fit quality and, when unfitted peaks remain, we promote the one with the smallest binding energy to an additional dimer candidate for the next fitting iteration. With this method, we are able to extract dimer binding energies even at larger $n$, where polyatomic contributions overlap densely.
The procedure deliberately minimizes false-positive dimer assignments at the cost of potentially overlooking weak dimer peaks.
As discussed in Sec.~\ref{sec:results}, Fig.~\ref{fig:overview_all_states} shows all dimer binding energies extracted from this fitting procedure, together with the corresponding energies predicted by our theoretical model. For a few peaks at $n\ge 43$ that were not unambiguously identified as dimers by the fitting procedure, we confirm their assignment as a dimer by comparison to the theoretical model, based on their clear correspondence to a calculated state.

\section{Calculation of phase shifts using a model potential}
\label{sec:appendixPS}
The scattering phase shifts are computed by solving 
the radial Schrödinger equation for an electron moving in the field of the neutral Yb atom with an interaction given by the model potential of Eq.~\ref{eq:model_potential}. 
To include the spin-orbit term, we must evaluate
   \begin{equation}
       B_{\alpha\alpha'}= \langle (LS)JM_J|\vec l_r\cdot\vec s_r|(L'S')J'M_J'\rangle,
   \end{equation}
   where we have dropped the $a$ subscripts. 
Eq.~7.1.6 of Edmonds~\cite{edmonds1996angular} shows how to evaluate the scalar product of two commuting tensor operators, giving
    \begin{equation}
    \begin{aligned}
       B_{\alpha\alpha'} &= \delta_{JJ'}\delta_{M_JM_J'}(-1)^{L'+S+J)}\begin{Bmatrix}J & S & L \\ 1 & L' & S'\end{Bmatrix}\\&\times\langle L||\vec l_r||L'\rangle \langle S||\vec s_r||S'\rangle\nonumber.
       \end{aligned}
       \end{equation}
       Here,
       \begin{equation}
       \begin{aligned}
       \langle L||\vec l_r||L'\rangle &=\delta_{LL'}\sqrt{(2L+1)(L+1)L}\nonumber
       \end{aligned}
       \end{equation}
       and
       \begin{equation}
       \begin{aligned}
       \langle S||\vec s_r||S'\rangle &= (-1)^{S'-\frac{1}{2}}\sqrt{\frac{3(2S+1)(2S'+1)}{2}}\begin{Bmatrix}S & 1 & S'\\ 1/2 & 0 & 1/2\end{Bmatrix} \nonumber
   \end{aligned}
       \end{equation}
   since we have $\vec l_r = \vec L$, and the total spin of Yb in its ground state is zero. 
   Evaluating these matrix elements gives
   \begin{equation}
       \begin{aligned}
       B_{(1,\frac{1}{2})\frac{1}{2}M_J,(1,\frac{1}{2})\frac{1}{2}M_J}&=-1\\
        B_{(1,\frac{1}{2})\frac{3}{2}M_J,(1,\frac{1}{2})\frac{3}{2}M_J}&=\frac{1}{2}.
   \end{aligned}
       \end{equation}
The two fine-structure components are diagonal in the $LS$ representation. 

The spin-orbit potential (Eq.~\ref{eq:so}) leads to an unphysical singularity $r^{-3}$ as $r\to0$. 
This was resolved in Ref.~\cite{bahrim2001boundary} by solving the Dirac equation at small distances to obtain the correct boundary condition at $r = r_0$ before propagating the standard Schrödinger equation out to larger internuclear distances. 
The Dirac equation is solved for a Hamiltonian involving only a pure Coulomb potential, which is the dominant term at very small $r$. The radial wave function in the $jj$-representation is
\begin{equation}
\begin{aligned}
u_{jl}(r) = \Bigl[&\, 1 - r\frac{\mathrm{d}f}{\mathrm{d}r}\frac{\mathrm{d}}{\mathrm{d}r}\frac{1}{r} - f\Bigl(\frac{\mathrm{d}^2}{\mathrm{d}r^2} - \frac{2}{r^2} \Bigr) \\
& + \frac{1}{r} \frac{\mathrm{d}f}{\mathrm{d}r}\Bigl(j(j+1)-\frac{11}{4}\Bigr) \, \Bigr] G_\kappa(r)
\end{aligned}
\end{equation}
where $\kappa$ is the Dirac relativistic quantum number $(l-j)(2j+1)$ and $f = 8c^2\cdot(1-\frac{-Zc}{2c^2r})^{-2}$. 
Here, 
$G_{\kappa}=(\kappa -s)J_{2s}(y) + \frac{y}{2}J_{2s+1}(y)$ is the solution to the Dirac equation at zero energy in the pure Coulomb potential.
The radial wave function $u_{jl}(r)$ is then transformed from $jj$ to $LS$ representations by
\begin{equation}
    u_{LSJ}(r) = (-1)^{1/2+L+J} \cdot \sqrt{2(2J+1)} 
    \begin{Bmatrix}
        L & J & 1/2 \\
        0 & 1/2 & J
    \end{Bmatrix},
\end{equation}
where it is already taken into account that the ground-state atom has zero spin and orbital angular momentum. 

\section{Effective range theory}\label{sec:appendixERT}
When the electron scattering energy is very low ($k \lesssim 0.01$),  the energy dependence of the $s$-wave scattering length $a_s(k)$ is governed by effective-range theory for a polarization potential~\cite{o1962low,o1964low}:
\begin{equation}
\begin{aligned}
\label{eq:effectiverange}
    a_s(k) = & a_s(0) + \frac{\pi \alpha}{3}k + \frac{4}{3}a_s(0)k^2{\rm{ln}}(1.23\sqrt{\alpha}k)  \\&+\bigg(\frac{R_e}{2} +\sqrt{\alpha}\frac{\pi}{3} -\alpha^{3/2}\frac{\pi}{3a_s(0)^2} \bigg) a_s(0)^2k^2.
\end{aligned}
\end{equation}
Before fitting the parameters of the model potential, it proved advantageous to treat the zero-energy scattering length $a_s(0)$, the polarizability $\alpha$, and the effective range parameter $R_e$ as fit parameters. 
Similarly, we used the Breit-Wigner form for a resonant phase shift to obtain a simple model for the $p_j$-wave phase shifts.
For $j=1/2$ we use
\begin{equation}
    \delta_{p_{1/2}}(k) = -\tan^{-1}\frac{\Gamma}{2E_{1/2}} + \tan^{-1} \frac{\Gamma}{2(E_{1/2} -\frac{k^2}{2})} ,
\end{equation}
with adjustable resonance position $E_{1/2}$ and width $\Gamma_{1/2}$.  
We adopt the same form for $j=3/2$, fixing $E_{3/2}=E_{1/2}+\SI{60}{\milli\eV}$ based on the fine-structure interval calculated in Ref.~\cite{dzuba1998low} and define $\Gamma_{3/2}=\Gamma_{1/2}$. 
In our fitting procedure, we first minimized the difference between theoretical and experimental spectra to fit these five parameters, and then obtained many different sets of parameters for the model potential which reproduced the low-energy scattering phase shifts (for $s$-wave) or the resonance position and width (for $p_{1/2}$-wave). We then proceeded to refine the fit using the model potential as discussed in the main text. 
The parameters used to generate the phase shifts displayed in 
Fig.~\ref{fig:phase_shifts} are: $\alpha = 141.203846, \, \lambda = 5.25602, \, A^{S} = 13.045043, \, \gamma^{S} = 25.476948, \, r_c^S = 3.755942, \, A^P = 28.394921, \, \gamma^P =25.505122, \, r_c^P = 2.296864$, where the superscript denotes if the parameter was used for $S$ or $P$-wave scattering. 

\section{Diagonalization method}
\label{sec:appendixDiag}
The large spin-orbit splitting of the $p$-wave interaction is essential to include for an accurate description of Yb ULRMs. 
This requires a more general form of the Fermi pseudopotential than the spin-independent case given in Eq.~\ref{eq:fermi}.
This was first presented for single-electron atoms in Ref.~\cite{eilesHamiltonian2017} and generalized to $LS$-coupled two-electron states in Ref.~\cite{wojciechowska2025ultralong}; here we reproduce the resulting matrix elements for our present system. 
The interaction matrix, described in the basis of Rydberg states $\ket{\alpha}$, is
\begin{equation}
    \underline{V} = 
\underline{A}\times\underline{U}\times\underline{A}^\dagger
\end{equation}
where 
\begin{equation}
    U_{\beta\beta'} = \delta_{\beta\beta'}\frac{(2L_a+1)^2}{2}\left[-\frac{\tan\delta_{^{2S_a+1}{L_a}_{J_a}}}{k^{2L_a+1}}\right]
\end{equation}
is a diagonal matrix in the spherical basis centered on the perturber, $\ket{\beta}=\ket{(L_aS_a)J_aM_{J_a}}$, and 
\begin{equation}
\begin{aligned}
A_{\alpha\beta}=  \sum
&C_{L_RM_{L_R},S_RM_{S_R}}^{J_RM_{J_R}}C_{l_rm_{l_r},l_cm_{l_c}}^{L_RM_{L_R}} C_{s_rm_{s_r},s_cm_{s_c}}^{S_RM_{S_R}}  \\
    &\cdot C_{L_aM_{L_a}, S_aM_{S_a}}^{J_aM_{J_a}} f_{L_a} Q^{L_a M_{L_a}}_{nl_rM_{L_a}}(R).
\end{aligned}
\end{equation}
effects the frame transformation into the basis of Rydberg states $\ket{\alpha}$. 
The sum ranges over all allowed values of $m$ in the subscripts of the Clebsch-Gordan coefficients. 
The first three Clebsch Gordan coefficients result from fully decoupling the Rydberg wave function, while the fourth results from uncoupling the scattering state. $f_L = \sqrt{4\pi/(2L+1)}$ and $Q$ is proportional to the radial wave function for $L_a=0$ and $L_a=1$, $M_{L_a}=\pm 1$ and proportional to its derivative with respect to $R$ for $L_a = 1$ and $M_{L_a}=0$~\cite{wojciechowska2025ultralong}. 
We note that the applied theory can be readily extended to include the hyperfine structure of the perturber~\cite{anderson2014photoassociation, eilesHamiltonian2017,eiles2023}. 

In practice, the diagonalization method's convergence issues (as discussed in the main text) make it inferior to the Green's function method for calculations with spectroscopic accuracy. However, as it is at present challenging to calculate electronic wave functions using the Green's function method, the diagonalization method has its utility when calculating transition dipole matrix elements or other wave function-based observables. 
Furthermore, the Green's function method numerically struggles to capture near-degenerate energy levels due to its reliance on a root-finding procedure, and it is not straightforward to add external fields. For these reasons we find that a combination of both numerical methods is valuable in ULRM studies.

\section{Generalizing the Green's function method to Yb}
\label{sec:appendixGF}
In this method, the Coulomb Green's function, whose closed form was first derived in Refs.~\cite{hostler1963coulomb}, is used to connect the scattering solution of the electronic wave function near the perturber with the Rydberg solution outside the perturber region.
This matching leads to a determinantal equation for the electronic energy. Details of this method are found in Ref.~\cite{eiles2023}; we give here only the appropriate changes to the equations in order to generalize the method to include a core electron within the $LS$ coupling scheme. 
We replace Eqs.~16 and 17 of~\cite{eiles2023}, which give the set of quantum numbers (except for $L$ in which the Coulomb Green's function is diagonal, with
     \begin{equation*}
       \begin{aligned}
		{\bf i} &\equiv {L,M_L, {s,m_s} f, m_f}\\
        \ket{\bf i} &\equiv \ket{LM_L,{(s_Rs_c)sm_s},(s_pI)fm_f } .
	\end{aligned}
    \end{equation*}
    Since the Green's function is diagonal in all quantum numbers in $i$ except $L$, we define ${\bf\bar{i}}$, consisting of all ${\bf i}$ quantum numbers except $L$.  
 We define  $\delta_{{\bf \bar{i},\bar{i}'}} \equiv \delta_{f,f'}\delta_{m_{f},m_{f'}}{\delta_{m_s,m_s'}\delta_{s,s'}}\delta_{M_{L},M_{L}'}$. 
 We additionally generalize the coupling coefficient $\cal S$ in Eq.~21 of Ref.~\cite{eiles2023} so that it describes the uncoupling of the total angular momentum of the Rydberg atom,
 	\begin{equation*}
       \begin{aligned}
			{\cal S}^{\ell,j}_{M_{L},M_{L'},{m_{s},m_{s'}}}&\equiv C_{lM_{L},{sm_{s}}}^{jM_{L}+m_{s}}C_{lM_{L'},{sm_{s'}}}^{jM_{L'}+m_{s'}}.
		\end{aligned}
    \end{equation*}
        Finally, 
        as with the shorthand index ${\bf i}$, here it is useful to define a second index, 
 \begin{align}\label{eq:ashort}
		{\boldsymbol \alpha} &\equiv {S,L,J,M_J,I,m_{I}, {m_c}},
    \end{align}
 along with the state 
 \begin{equation}
   \ket{\boldsymbol{\alpha}} \equiv \ket{[(s_Rs_p)SL]JM_J,Im_I}{\ket{s_cm_c}},
 \end{equation}
	incorporating all of the degrees of freedom of the perturber spins and atom-electron scattering complex. 
    The spin recoupling matrix elements $\langle {\bf i}|\boldsymbol\alpha \rangle$ are
    	\begin{equation*}\label{eq:recoup}
        \begin{aligned}
		\mathcal{A}_{i\alpha}\equiv \langle{\bf i}  | \boldsymbol{\alpha}  \rangle =\langle \boldsymbol{\alpha} | {\bf i} \rangle &=  
		\delta_{{L_i},{L_\alpha}} \Big[\sum_{{m_{p}},{M_S},{m_R}}
		C_{S M_{S},L_\alpha M_{L}}^{J M_{J}}  \\&\times C_{s_R m_{R},s_pm_{p}}^{S M_{S}} C_{s_p m_{p},I m_{I}}^{f m_{f}} {C_{s_Rm_R,s_cm_c}^{sm_s}}\Big].
	\end{aligned}
    \end{equation*}
A similar set of modifications could be employed to generalize the treatment of Ref.~\cite{eiles2023} to $jj$-coupled Rydberg systems. 
\section{Calculation of vibrational levels}
\label{sec:appendixBS}
We obtain the vibrational spectrum upon solving the nuclear Schrödinger equation in the Born-Oppenheimer approximation,
\begin{equation}
\label{eq:vibstates}
  0=  \left(-\frac{1}{2M_r}\frac{\mathrm{d}^2}{\mathrm{d}R^2} + U_\alpha(R) - E_v^\alpha\right)\chi_v^\alpha(R).
\end{equation}
Because of the large reduced mass $M_r=173.938/2$ and the long bond lengths, ranging from $\num{700}$ to $\num{2500}\,a_0$, the rotational splitting is on the order of only a few tens of $\si{\kilo\hertz}$.
We therefore focus on the ground rotational state. 
To accurately treat the fact that most of the excited molecular states are resonances whose widths are determined by the strength of quantum reflection off of the steep drop in the potential near the crossing with the $p_{1/2}$ butterfly state, we use the method of Siegert pseudostates~\cite{tolstikhin1997siegert} to solve Eq.~\ref{eq:vibstates} subject to ingoing-wave boundary conditions. 
This yields vibrational resonance positions and widths; we ignore vibrational resonances with widths exceeding $\SI{15}{\mega\hertz}$ as too broad to be experimentally resolved. 
Non-adiabatic effects are neglected in this work, as we estimate the error due to them to be on the level of a few percent of the total binding energy~\cite{durst_nonadiabatic_2025}.

\bibliography{references}
\end{document}